\documentclass{jpp}
\usepackage{graphicx}
\usepackage{epstopdf, epsfig}
\usepackage{hyperref}
\usepackage{amsmath,amsfonts,amssymb}
\usepackage{subfigure}

\newcommand{\elsasser}{Els\"asser}
\newcommand{\alfven}{Alfv\'en}

\shorttitle{Waveframe variables for MHD}
\shortauthor{Van Doorsselaere et al.}

\title{The magnetohydrodynamic equations in terms of waveframe variables}

\author{T. Van Doorsselaere\aff{1}\corresp{\email{tom.vandoorsselaere@kuleuven.be}}, N. Magyar\aff{1},  M.~V. Sieyra\aff{2} and M. Goossens\aff{1}}

\affiliation{\aff{1}Centre for mathematical Plasma Astrophysics, Department of Mathematics, KU~Leuven, Celestijnenlaan 200B, B-3001 Leuven, Belgium
\aff{2}D\'epartement d'Astrophysique/AIM, CEA/IRFU, CNRS/INSU, Universit\'e Paris-Saclay, Universit\'e de Paris, F-91191,
Gif-sur-Yvette, France}

\begin{document}

\maketitle

\begin{abstract}
Generalising the \elsasser{} variables, we introduce the $Q$-variables. These are more flexible than the \elsasser{} variables,
because they also allow to track waves with phase speeds different than the \alfven{} speed. We rewrite the MHD equations with these $Q$-variables. We
consider also the linearised version of the resulting MHD equations in a uniform plasma, and recover the classical \alfven{} waves, but also
separate the fast and slow magnetosonic waves in upward and downward propagating waves. Moreover, we show that the $Q$-variables may also track the
upward and downward propagating surface \alfven{} waves in a non-uniform plasma, displaying the power of our generalisation. In the end, we lay the
mathematical framework for driving solar wind models with a multitude of wave drivers.
\end{abstract}

\keywords{to be added in editing}

\section{Introduction} \label{sec:intro}
		The \elsasser{} variables \citep{elsasser1950} are expressed as
		\begin{equation}
			\vec{Z}^\pm=\vec{V}\pm\vec{V}_\mathrm{A},
		\end{equation}
		where $\vec{V}$ is the speed of the plasma and $\vec{V}_\mathrm{A}=\vec{B}/\mu\rho$ is the vectorial \alfven{} speed expressed in terms of the
magnetic field $\vec{B}$, density $\rho$ and magnetic permeability $\mu$. In MHD, these \elsasser{} variables play a unique role which conveniently
corresponds to \alfven{} waves. A single \alfven{} wave may be expressed through a single \elsasser{} variable.\\
		Because of this convenient property and the prevalence of \alfven{} wave turbulence in the solar wind, the \elsasser{} variables have been
used numerous times in the description of the plasma in the solar wind
\citep[e.g.][]{dobrowolny1980,velli1989,marsch1989,zhou1989,tu1989,grappin1990,bruno2013}. With the \elsasser{} variables, it is straightforward to
show that \alfven{} wave turbulence exists because of the interaction of counterpropagating \alfven{} waves \citep{bruno2013} in incompressible MHD.
Given the great success of the \elsasser{} variables, \citet{marsch1987} have even gone so far as to rewrite the entire set of MHD equations in terms
of the independent variables, comprising of the  \elsasser{} variables and the density. In that paper, it is clear that the entire machinery of MHD
waves can be recovered for this set of equations in terms of \elsasser{} variables and density. This set of equations offers the possibility to study
the evolution of MHD waves through the \elsasser{} variables. The caveat is that the \elsasser{} variables are really only well suited to model
\alfven{} waves.
		\par
		However, for other waves, the \elsasser{} variables are less well suited, because other MHD waves necessarily consist of a combination of both
\elsasser{} variables. For example, \citet{magyar2019a} show that this is particularly true for slow and fast magnetosonic waves in a homogeneous
plasma. But this statement also holds for most waves in a non-uniform plasma. \citet{ismayilli2022} calculated the \elsasser{} variables for surface
\alfven{} waves on a discontinuous interface between two homogeneous plasmas, and clearly show that both \elsasser{} components are non-zero for this
surface \alfven{} wave. Moreover, the \elsasser{} variables are no longer uniquely associated with upward or downward propagation. For instance, an
upward propagating kink wave in a cylindrical plasma has both \elsasser{} variables co-propagating along the magnetic field \citep{vd2020}. Their
continuous interaction would lead to an efficient formation of turbulence, and this turbulence from a unidirectional transverse wave is called
uniturbulence \citep{magyar2017}. To study the non-linear evolution of such waves in inhomogeneous plasmas, a more general approach than \elsasser{}
variables is needed. \par
		In direct measurements in the solar wind, it has been found many times that the magnetic field fluctuations and the velocity fluctuations 
are highly correlated, showing that they are highly Alfv\'enic \citep{bavassano2000}. This is expressed through the \alfven{} ratio 
$r_\mathrm{A}$, which is the ratio of the kinetic energy and the magnetic energy, which is found to be close to 1 close to the Sun. However, it 
has also been found in solar wind data that the slope of the correlation between the magnetic field fluctuations and the velocity fluctuations is not 
always 1 \citep{marsch1993}. This is potentially because of the presence of other wave modes than \alfven{} waves. Thus, also observationally, there 
is a need for a generalisation of the \elsasser{} variables. 
		\par
		Here we consider a generalisation of the \elsasser{} variables by considering them as co-moving with the wave, using the phase speed as a
parameter. We call these the $Q$-variables. However, the push for a generalisation of \elsasser{} variables is embraced in the wider community. For
example, \citet{galtier2023} has considered so-called canonical variables. With these canonical variables, he described successfully the interaction
and cascade of fast mode waves. Thus, it seems that more general \elsasser{} variables are possible, and this should be a research question that is
actively pursued, given the tremendous impact of the \elsasser{} variables.

		\section{Results}
		\subsection{The MHD equations written in terms of $Q$-variables}
		In what follows, we introduce a new parameter $\alpha$, which describes the wave phase speed, for a general wave. We then introduce the
$Q$-variables by
		\begin{equation}
			\vec{Q}^\pm=\vec{V}\pm \alpha \vec{B},
		\end{equation}
		where it is clear that the limit $\alpha=1/\sqrt{\mu \rho}$ recovers the special case of \elsasser{} variables. Taking this limit thus always
allows to check our equations against the relevant equations in \citet{marsch1987}.\par

		We start from the same set of ideal MHD equations as \citet{marsch1987} do. They read
		\begin{align}
			\frac{\partial \vec{V} }{\partial t}+\vec{V}\cdot \nabla \vec{V} &= -\frac{1}{\rho}\nabla P_T + \frac{1}{\mu\rho}\vec{B}\cdot\nabla
\vec{B}, \label{eq:momeq}\\*
			\frac{\partial \ln \rho}{\partial t}+\vec{V}\cdot \nabla \ln{\rho}& =-\nabla\cdot\vec{V}, \label{eq:conteq}\\*
			\frac{\partial \vec{B}}{\partial t}&=-\vec{B}\nabla\cdot\vec{V}-\vec{V}\cdot\nabla\vec{B} + \vec{B}\cdot\nabla\vec{V}, \label{eq:indeq}\\*
			\nabla\cdot\vec{B} &=0, \label{eq:soleq}
		\end{align}
		where the total pressure is defined as $P_\mathrm{T}=p+\frac{1}{2}\rho V_\mathrm{A}^2$, using the gas pressure $p$ and vectorial \alfven{}
speed $\vec{V}_\mathrm{A}=\vec{B}/\sqrt{\mu \rho}$. They are complimented with an adiabatic assumption for the energy equation
		\begin{equation}
			p=p(\rho)=p_0(\rho/\rho_0)^\gamma,
		\end{equation}
		where $\gamma$ is the adiabatic exponent.

		\subsubsection{Solenoidal constraint}
		Let us first consider the solenoidal constraint (Eq.~\ref{eq:soleq}). We rewrite it in terms of $Q$-variables, through the expression of
$\vec{B}$ in terms of $\vec{Q}^\pm$:
		\begin{equation}
			\vec{B}=\frac{1}{2\alpha}(\vec{Q}^+-\vec{Q}^-). \label{eq:B}
		\end{equation}
		Inserting that into Eq.~\ref{eq:soleq} allows to write
		\begin{equation}
			0=\frac{1}{2\alpha}\nabla\cdot(\vec{Q}^+-\vec{Q}^-)-\frac{1}{2\alpha}(\vec{Q}^+-\vec{Q}^-)\cdot\nabla\ln{\alpha},
		\end{equation}
		or, after simplification,
		\begin{equation}
			0=\nabla\cdot(\vec{Q}^+-\vec{Q}^-)-(\vec{Q}^+-\vec{Q}^-)\cdot\nabla\ln{\alpha}. \label{eq:divb}
		\end{equation}
		Considering the limiting case of $\alpha^2=1/\mu\rho, \vec{Q}^\pm=\vec{Z}^\pm, \vec{Z}^+-\vec{Z}^-=\vec{V}_\mathrm{A}$, we recover Eq.~(10) of
\citet{marsch1987}.

		\subsubsection{Conservation of mass}
		Next, we rewrite the conservation of mass (Eq.~\ref{eq:conteq}). We use it for finding an expression for $\frac{D^\pm}{Dt} (\ln{\rho})$, where
$D^\pm/Dt=\partial/\partial t+\vec{Q}^\pm\cdot\nabla$ is the derivative co-moving with the wave, in the so-called waveframe. We find
		\begin{align}
			\frac{D^\pm}{Dt} (\ln{\rho}) &= \frac{\partial\ln{\rho}}{\partial t}+\vec{Q}^\pm\cdot\nabla \ln{\rho}, \\*
			&= \frac{\partial\ln{\rho}}{\partial t} + \vec{V}\cdot\nabla\ln{\rho} \pm \alpha\vec{B}\cdot\nabla\ln{\rho},\\*
			&= -\nabla\cdot\vec{V}\pm \alpha\vec{B}\cdot\nabla\ln{\rho},
		\end{align}
		where the continuity equation (Eq.~\ref{eq:conteq}) was used in the last equation. To this last equation, we add in the righthand side
($\mp\times$ Eq.~\ref{eq:divb}) to find
		\begin{align}
			\frac{D^\pm}{Dt} (\ln{\rho}) &= -\frac{1}{2}\nabla\cdot(\vec{Q}^++\vec{Q}^-)\mp \nabla\cdot(\vec{Q}^+-\vec{Q}^-)\pm
\frac{\vec{Q}^+-\vec{Q}^-}{2}\cdot\nabla\ln{\rho\alpha^2}, \label{eq:clos1} \\*
			&= -\frac{1}{2}\nabla\cdot(3\vec{Q}^\pm-\vec{Q}^\mp)\pm \frac{\vec{Q}^+-\vec{Q}^-}{2}\cdot\nabla\ln{\rho\alpha^2}, \label{eq:clos2}
		\end{align}
		where we have used the expressions for $\vec{V}$ in terms of $\vec{Q}^\pm$ in the equations:
		\begin{equation}
			\vec{V}=\frac{1}{2}(\vec{Q}^++\vec{Q}^-). \label{eq:V}
		\end{equation}
		When the limit of $\alpha^2\to 1/\sqrt{\mu\rho}$ is considered, the last term of Eq.~\ref{eq:clos2} cancels out and Eq.~16 of
\citet{marsch1987} is readily recovered.
		\par

		\subsubsection{Momentum equation}
		Now we turn to the momentum equation and the induction equation (Eqs.~\ref{eq:momeq} and \ref{eq:indeq}), which form the key equation (17) of
\citet{marsch1987}. Following their lead, we add (Eq.~\ref{eq:momeq})$\pm\alpha$(Eq.~\ref{eq:indeq}). In the first step, we use the expansion of
$\vec{Q}^\mp\cdot\nabla\vec{Q}^\pm$ as
		\begin{equation}
			\vec{Q}^\mp\cdot\nabla\vec{Q}^\pm=\vec{V}\cdot\nabla\vec{V}\mp \alpha \vec{B}\cdot\nabla\vec{V}\pm \alpha \vec{V}\cdot\nabla \vec{B}
-\alpha^2\vec{B}\cdot\nabla\vec{B}\pm\vec{B}\vec{Q}^\mp \cdot\nabla \alpha,
		\end{equation}
		where we have used the vector identity $\vec{C}\cdot\nabla(f\vec{D})=f\vec{C}\cdot\nabla\vec{D}+\vec{D}(\vec{C}\cdot\nabla f)$ for any vector
fields $\vec{C}$ and $\vec{D}$ and scalar field $f$. We also define the parameter
		\begin{equation}
			\Delta \alpha^2=\alpha^2-\frac{1}{\mu\rho}.
		\end{equation}
		The $\Delta \alpha^2$ parameter expresses how far a wave's phase speed is from the \alfven{} speed. Since a wave can be slower or faster than 
the \alfven{} speed, the $\Delta \alpha^2$ parameter may be positive or negative, despite the square! The square in the notation is kept for 
dimensional purposes to keep $\Delta \alpha$ in the same units as $\alpha$. In the limit of $\alpha=\frac{1}{\sqrt{\mu\rho}}$,
the parameter $\Delta \alpha^2$ will turn to 0: $\Delta \alpha^2=0$ and $\vec{Q}^\pm=\vec{Z}^\pm$ turns into the classical \elsasser{} variable. Here 
it is also useful to point
out that it will be convenient to use expressions with $\rho\alpha^2$, which are constant in this limit. \\
		With the above expressions, we obtain from (Eq.~\ref{eq:momeq})$\pm\alpha$(Eq.~\ref{eq:indeq}) the result
		\begin{equation}
			\frac{\partial \vec{Q}^\pm}{\partial t}\mp \vec{B}\frac{\partial \alpha}{\partial t}=
-\vec{Q}^\mp\cdot\nabla\vec{Q}^\pm\pm\vec{B}\vec{Q}^\mp\cdot\nabla \alpha-\Delta \alpha^2\vec{B}\cdot\nabla\vec{B}-\frac{1}{\rho}\nabla P_\mathrm{T} 
\mp
\alpha \vec{B}\nabla\cdot\vec{V}.
		\end{equation}
		The first two terms on the righthand side group with the lefthand side to form the co-moving derivative:
		\begin{equation}
			\frac{D^\mp }{Dt}\vec{Q}^\pm\mp\vec{B}\frac{D^\mp }{Dt}\alpha=-\frac{1}{\rho}\nabla P_\mathrm{T} -\Delta \alpha^2\vec{B}\cdot\nabla\vec{B} 
\mp
\alpha \vec{B}\nabla\cdot\vec{V}. \label{eq:adv}
		\end{equation}\\
		We now find an expression for the terms in the righthand side. For the total pressure term, we find
		\begin{align}
			\frac{1}{\rho}\nabla P_\mathrm{T}&=\frac{1}{\rho}\nabla \left(p+\frac{B^2}{2\mu}\right),\\*
			&= v_\mathrm{s}^2 \nabla \ln{\rho} + \frac{1}{8\alpha^2} (\alpha^2-\Delta \alpha^2)\nabla (\vec{Q}^+-\vec{Q}^-)^2 +
\frac{1}{8}(\alpha^2-\Delta \alpha^2)(\vec{Q}^+-\vec{Q}^-)^2\nabla \left(\frac{1}{\alpha^2}\right),\\*
			&= v_\mathrm{s}^2 \nabla \ln{\rho} + \frac{1}{8} \left(1-\frac{\Delta \alpha^2}{\alpha^2}\right)\nabla (\vec{Q}^+-\vec{Q}^-)^2 -
\frac{1}{4}\left(1-\frac{\Delta \alpha^2}{\alpha^2}\right)(\vec{Q}^+-\vec{Q}^-)^2\nabla \ln{\alpha},
		\end{align}
		where we have used the adiabatic relationship of $p(\rho)$ which introduces the expression for the sound speed
$v_\mathrm{s}=\sqrt{\frac{\gamma p}{\rho}}$. \\
		The second term in the righthand side of Eq.~\ref{eq:adv} can be rewritten with the expression for $\vec{B}$ in terms of $\vec{Q}^\pm$ as
		\begin{equation}
			-\Delta \alpha^2\vec{B}\cdot\nabla\vec{B} = -\frac{1}{4}\frac{\Delta \alpha^2}{\alpha^2} (\vec{Q}^+-\vec{Q}^-)\cdot \nabla 
(\vec{Q}^+-\vec{Q}^-) +
\frac{1}{4}\frac{\Delta \alpha^2}{\alpha^2} (\vec{Q}^+-\vec{Q}^-) \left( (\vec{Q}^+-\vec{Q}^-)\cdot\nabla \ln{\alpha}\right).
		\end{equation}\\
		The third term in the righthand side of Eq.~\ref{eq:adv} should be handled through the modified version of the continuity relation
Eq.~\ref{eq:clos1}. From that equation, we have that
		\begin{multline}
			\mp\alpha \vec{B}\nabla\cdot\vec{V} = \pm \left(\frac{\vec{Q}^+-\vec{Q}^-}{2}\right) \frac{D^\mp}{Dt} (\ln{\rho}) -
\left(\frac{\vec{Q}^+-\vec{Q}^-}{2}\right) \nabla \cdot (\vec{Q}^+-\vec{Q}^-)\\* + \left(\frac{\vec{Q}^+-\vec{Q}^-}{2}\right)
\left(\left(\frac{\vec{Q}^+-\vec{Q}^-}{2}\right)\cdot\nabla \ln{\rho\alpha^2}\right).
		\end{multline}
		Substituting everything in Eq.~\ref{eq:adv}, we now have
		\begin{multline}
			\frac{D^\mp }{Dt}\vec{Q}^\pm\mp\left(\frac{\vec{Q}^+-\vec{Q}^-}{2}\right)\frac{D^\mp }{Dt}\ln{\alpha}=-v_\mathrm{s}^2 \nabla \ln{\rho} -
\frac{1}{8} \left(1-\frac{\Delta \alpha^2}{\alpha^2}\right)\nabla (\vec{Q}^+-\vec{Q}^-)^2 \\ +
\frac{1}{4}\left(1-\frac{\Delta \alpha^2}{\alpha^2}\right)(\vec{Q}^+-\vec{Q}^-)^2\nabla \ln{\alpha} -\frac{1}{4}\frac{\Delta \alpha^2}{\alpha^2}
(\vec{Q}^+-\vec{Q}^-)\cdot \nabla (\vec{Q}^+-\vec{Q}^-) \\ + \frac{1}{4}\frac{\Delta \alpha^2}{\alpha^2} (\vec{Q}^+-\vec{Q}^-) \left(
(\vec{Q}^+-\vec{Q}^-)\cdot\nabla \ln{\alpha}\right) \pm \left(\frac{\vec{Q}^+-\vec{Q}^-}{2}\right) \frac{D^\mp}{Dt} (\ln{\rho})\\* -
\left(\frac{\vec{Q}^+-\vec{Q}^-}{2}\right) \nabla \cdot (\vec{Q}^+-\vec{Q}^-) + \left(\frac{\vec{Q}^+-\vec{Q}^-}{2}\right)
\left(\left(\frac{\vec{Q}^+-\vec{Q}^-}{2}\right)\cdot\nabla \ln{\rho\alpha^2}\right).
		\end{multline}
		After moving the RHS convective derivative to the LHS and subsequently adding ($\pm \left(\frac{\vec{Q}^+-\vec{Q}^-}{4}\right)\times$
(Eq.~\ref{eq:clos2})) and using Eq.~\ref{eq:divb}, we obtain the final result
		\begin{multline}
			\frac{D^\mp }{Dt}\vec{Q}^\pm\mp\left(\frac{\vec{Q}^+-\vec{Q}^-}{4}\right)\frac{D^\mp }{Dt}\ln{\rho\alpha^2}=-v_\mathrm{s}^2 \nabla
\ln{\rho} - \frac{1}{8} \left(1-\frac{\Delta \alpha^2}{\alpha^2}\right)\nabla (\vec{Q}^+-\vec{Q}^-)^2 \\+
\frac{1}{4}\left(1-\frac{\Delta \alpha^2}{\alpha^2}\right)(\vec{Q}^+-\vec{Q}^-)^2\nabla \ln{\alpha}  -\frac{1}{4}\frac{\Delta \alpha^2}{\alpha^2}
(\vec{Q}^+-\vec{Q}^-)\cdot \nabla (\vec{Q}^+-\vec{Q}^-) \\ + \frac{1}{4}\frac{\Delta \alpha^2}{\alpha^2} (\vec{Q}^+-\vec{Q}^-) 
\nabla\cdot(\vec{Q}^+-\vec{Q}^-)
 \mp  \left(\frac{\vec{Q}^+-\vec{Q}^-}{8}\right) \nabla \cdot (3\vec{Q}^\pm-\vec{Q}^\mp) \\+ \left(\frac{\vec{Q}^+-\vec{Q}^-}{4}\right)
\left(\left(\frac{\vec{Q}^+-\vec{Q}^-}{2}\right)\cdot\nabla \ln{\rho\alpha^2}\right). \label{eq:qmhd}
		\end{multline}
		Taking the limit of $\alpha=\sqrt{1/\mu\rho}$ allows us to confirm that this equation converges in that case to Eq.~(17) of
\citet{marsch1987}. The last term in the LHS and the last three terms in the RHS are terms parallel to the magnetic field. We remind the reader that
these equations are valid for any choice of $\alpha$ (satisfying basic dimensional arguments).

		\subsection{Linearised $Q$-equations}\label{sec:linear}
		In a first attempt to better understand the $Q$-variables and the role that $\alpha$ plays in the MHD equations, we shall linearise the MHD
equations (Eqs.~\ref{eq:divb}, \ref{eq:clos2}, \ref{eq:qmhd}) around a uniform equilibrium. We take $\rho=\rho_0+\delta \rho,
\vec{B}=B_0\vec{e_z}+\vec{\delta B}, \vec{V}=\vec{V_0}+\vec{\delta V}, \vec{Q}^\pm=\vec{Q}_0^\pm+\vec{\delta Q}^\pm$, where quantities with subscript
0 are constant equilibrium quantities, and $\delta$ indicate Eulerian perturbations (where we have used the Chandrasekhar notation for such). The
Cartesian coordinate system $(x,y,z)$ is aligned with the magnetic field in the $z$-direction. We have not linearised $\alpha$, because we shall show
later that it is proportional to the phase speed of the wave. Moreover, a linearisation of $\alpha$ would result in terms rewritten from 
$\delta \rho$ and
other physical parameters, and consequently the equation for the linearised $\alpha$ would be linearly dependent on the previous equations. \par
		Adopting a similar notation as \citet{marsch1987}, we have
		\begin{align}
			\nabla \ln{\rho} &= \nabla \ln{(\rho_0(1+\frac{\delta \rho}{\rho_0}))} = \nabla \frac{\delta \rho}{\rho_0} \equiv \nabla \delta R, \\ 
			\nabla \ln{\rho \alpha^2} &= \nabla \ln{\rho} + \nabla \ln{\alpha^2}= \nabla \delta R.
		\end{align}
		We have utilised that the background variables are uniform, and that $\alpha$ does not need to be linearised. We have also
rejected any terms higher than the first order in perturbations and defined the quantity $\delta R$. Additionally we linearise
the co-moving advective derivative $D^\pm/Dt$ as
		\begin{equation}
			\frac{D^\pm}{Dt}=\frac{\partial}{\partial t} + \vec{Q}^\pm_0\cdot \nabla + \vec{\delta Q}^\pm\cdot \nabla \equiv \frac{d^\pm}{dt} +
\vec{\delta Q}^\pm\cdot \nabla,
		\end{equation}
		where the notation of \citet{marsch1987} was once again used to define $d^\pm/dt$. Note also that the last term always results in 0 when
operating on equilibrium quantities, given their assumed homogeneity. Action of the last term on linear quantities result in a 2nd order
contribution, which is neglected. With this notation, the MHD equations are rewritten as
		\begin{align}
			\frac{d^\mp}{dt} \vec{\delta Q}^\pm & \mp \left(\frac{\vec{Q}^+_0-\vec{Q}^-_0}{4}\right) \frac{d^\mp}{dt}\delta R = -v_\mathrm{s0}^2
\nabla \delta R - \frac{1}{4}\left(1-\frac{\Delta \alpha^2}{\alpha^2}\right) \nabla \left((\vec{\delta Q}^+-\vec{\delta
Q}^-)\cdot(\vec{Q}^+_0-\vec{Q}^-_0)\right) \nonumber \\*
			&  - \frac{1}{4} \frac{\Delta \alpha^2}{\alpha^2} (\vec{Q}^+_0-\vec{Q}^-_0) \cdot \nabla (\vec{\delta Q}^+-\vec{\delta Q}^-) + \frac{1}{4}
\frac{\Delta \alpha^2}{\alpha^2} (\vec{Q}^+_0-\vec{Q}^-_0) \nabla \cdot  (\vec{\delta Q}^+-\vec{\delta Q}^-) \nonumber \\*
			& \mp \left(\frac{\vec{Q}^+_0-\vec{Q}^-_0}{8}\right) \nabla \cdot (3\vec{\delta Q}^\pm-\vec{\delta Q}^\mp) +
\left(\frac{\vec{Q}^+_0-\vec{Q}^-_0}{4}\right)\left( \left(\frac{\vec{Q}^+_0-\vec{Q}^-_0}{2}\right)\cdot \nabla \delta R\right), \label{eq:qeq1} \\*
			\frac{d^\pm}{dt} \delta R &= -\frac{1}{2} \nabla \cdot (3\vec{\delta Q}^\pm-\vec{\delta Q}^\mp) \pm
\left(\frac{\vec{Q}^+_0-\vec{Q}^-_0}{2}\right)\cdot \nabla \delta R, \label{eq:qeq2} \\*
			0 &= \nabla \cdot  (\vec{\delta Q}^+-\vec{\delta Q}^-). \label{eq:qeq3}
		\end{align}

		Given the homogeneity, the linear wave solutions may be written with the plane wave notation $\exp{(i\vec{k}\cdot\vec{x}-i\omega t)}$, where
we choose the $x$-axis to be in the $\vec{k}-\vec{B}_0$-plane resulting in $k_y\equiv 0$. For the plane waves, the co-moving derivative is rewritten
as $d^\pm/dt=-i(\omega - \vec{k}\cdot \vec{Q}^\pm_0)\equiv -i\omega^\pm$, where we have yet again used the notation of \citet{marsch1987}. With these
notations, we can split the $Q$-equations (Eqs.~\ref{eq:qeq1}-\ref{eq:qeq2}) in its components:
		\begin{align}
			-\omega^\mp\delta R &= -\frac{1}{2} k_x (3 \delta Q^\mp_x-\delta Q^\pm_x)-\frac{1}{2} k_z (3\delta Q^\mp_z-\delta Q^\pm_z) \mp \alpha
B_0k_z \delta R, \label{eq:Lr}\\*
			-\omega^\mp\delta Q^\pm_x &= -v_\mathrm{s0}^2 k_x \delta R - \frac{1}{2}\left(1-\frac{\Delta \alpha^2}{\alpha^2}\right) \alpha B_0 k_x 
(\delta
Q^+_z-\delta Q^-_z) - \frac{1}{2} \frac{\Delta \alpha^2}{\alpha^2} \alpha B_0k_z (\delta Q^+_x-\delta Q^-_x), \label{eq:Lx} \\*
			-\omega^\mp \delta Q^\pm_y &= - \frac{1}{2} \frac{\Delta \alpha^2}{\alpha^2} \alpha B_0k_z (\delta Q^+_y-\delta Q^-_y), \label{eq:Ly} \\*
			-\omega^\mp \delta Q^\pm_z & \pm \frac{1}{2}\alpha B_0 \omega^\mp \delta R = -v_\mathrm{s0}^2 k_z \delta R -
\frac{1}{2}\left(1-\frac{\Delta \alpha^2}{\alpha^2}\right) \alpha B_0 k_z (\delta Q^+_z-\delta Q^-_z)  \nonumber \\*
			& + \frac{1}{2} \frac{\Delta \alpha^2}{\alpha^2} \alpha B_0k_x (\delta Q^+_x-\delta Q^-_x) \mp \frac{1}{4} \alpha B_0 \left(k_x (3 \delta
Q^\pm_x-\delta Q^\mp_x) + k_z (3\delta Q^\pm_z-\delta Q^\mp_z)\right)\nonumber \\ & + \frac{1}{2}\alpha^2B_0^2 k_z \delta R, \label{eq:Lz}
		\end{align}
		which form a system of 7 equations for 7 unknowns. It has eigenvalue $\omega$. Remember, in these equations, $\alpha$ can still be chosen
freely!

		\subsubsection{\alfven{} waves}

		As expected, the $y$-component (Eq.~\ref{eq:Ly}) is separated from the other equations. This equation is rewritten in the following system
		\begin{align}
			(\omega - \vec{k}\cdot\vec{Q}^-_0)\delta Q^+_y &= \frac{1}{2} \frac{\Delta \alpha^2}{\alpha^2} \alpha B_0k_z (\delta Q^+_y-\delta
Q^-_y),
\label{eq:alfplus} \\*
			(\omega - \vec{k}\cdot\vec{Q}^+_0)\delta Q^-_y &= \frac{1}{2} \frac{\Delta \alpha^2}{\alpha^2} \alpha B_0k_z (\delta Q^+_y-\delta
Q^-_y),
\label{eq:alfmin}
		\end{align}
		resulting in a dispersion relation
		\begin{equation}
			\omega^2 - \vec{k}\cdot(\vec{Q}_0^++\vec{Q}_0^-)\omega +
(\vec{k}\cdot\vec{Q}^+_0)(\vec{k}\cdot\vec{Q}^-_0) + \frac{1}{2} \vec{k}\cdot(\vec{Q}_0^+-\vec{Q}_0^-)\frac{\Delta \alpha^2}{\alpha}B_0k_z=0,
		\end{equation}
		with solutions
		\begin{align}
			\omega & = \vec{k}\cdot\vec{V}_0 \pm \sqrt{( \vec{k}\cdot\vec{V}_0)^2 -
(\vec{k}\cdot\vec{Q}^+_0)(\vec{k}\cdot\vec{Q}^-_0) - \Delta \alpha^2 k_z^2 B_0^2 }\\*
			& = \vec{k}\cdot\vec{V}_0 \pm \sqrt{ \left(\frac{\vec{k}}{2}\cdot(\vec{Q}_0^+-\vec{Q}_0^-)\right)^2 - \Delta \alpha^2 k_z^2 B_0^2} \\*
			&= \vec{k}\cdot\vec{V}_0 \pm k_zB_0 \sqrt{\alpha^2-\Delta \alpha^2}\\*
			&= \vec{k}\cdot\vec{V}_0 \pm \frac{k_zB_0}{\sqrt{\mu\rho_0}},
		\end{align}
		which nicely converges to the well-known \alfven{} wave solution $\omega=\vec{k}\cdot(\vec{V}_0\pm\vec{V}_\mathrm{A})=\vec{k}\cdot
\vec{Z}^\pm_0$.

		This subsection also points us in the direction of the meaning and importance of the $\alpha$ parameter. If we would change variables to the
co-moving frame (co-moving with $\vec{Q}^\pm_0$), then that frame would require that either $\omega^\pm=0$ separately. Implementing these conditions
in Eq.~\ref{eq:alfplus} and \ref{eq:alfmin}, leads to the (single) condition
		\begin{equation}
			\frac{1}{2} \frac{\Delta \alpha^2}{\alpha^2} \alpha B_0k_z (\delta Q^+_y-\delta Q^-_y)=0.
		\end{equation}
		From this condition, we obtain that $(\delta Q^+_y-\delta Q^-_y)\equiv 0$ or that $\Delta \alpha^2\equiv 0$. The former condition would lead 
to $\delta
Q^\pm_y\equiv 0$ through the companion equation (e.g. Eq~\ref{eq:alfmin} for $\omega^-=0$), which tells us that there is no physical solution with
non-zero amplitude. The latter condition $\Delta \alpha^2=0$ leads to the well-known solution $\alpha^2=1/\mu\rho_0$, which is equivalent to the 
limit where the
$Q$-variables coincide with the \elsasser{} variables. This thus shows that the \elsasser{} variables are the only co-propagating waveframe variables
in which the \alfven{} waves have a non-zero amplitude. It shows that $\alpha$ should be chosen according to the phase speed, through the solution of
$\omega^\pm=0$:
		\begin{equation}
			0=\omega^\pm = \omega - \vec{k}\cdot \vec{Q}^\pm_0 = \omega - \vec{k}\cdot \vec{V}_0 \mp \alpha \vec{k}\cdot\vec{B}_0,
		\end{equation}
		resulting in an expression for $\alpha$:
		\begin{equation}
			\alpha = \pm \frac{\omega - \vec{k}\cdot \vec{V}_0}{\vec{k}\cdot\vec{B}_0}. \label{eq:alpha}
		\end{equation}
		The reader is cautioned to be careful with this expression, given that the expression diverges if $k\to 0$ or perpendicular $\vec{k}$ and
$\vec{B}_0$.

		\subsubsection{Magnetoacoustic waves}

		Let us now investigate magnetoacoustic waves as they appear in terms of $Q$-variables. For the specific geometry chosen without loss of
generality in Section~\ref{sec:linear}, linear magnetoacoustic modes perturb the $Q$-variables in the $x-z$ plane, and density. The system of
equations to be solved for magnetoacoustic modes is composed of Eqs.~\ref{eq:Lr}-\ref{eq:Lz}, except Eqs.~\ref{eq:Ly} which were treated in the
previous subsection, yielding \alfven{} waves. The dispersion relation is given by the determinant of this system of 5 equations for 5 unknowns.
However,
it turns out that in this system there are only 4 independent equations, Eq.~\ref{eq:Lr} being linearly dependent on the other equations. Instead, we
use the linearized solenoidal constraint (Eq.~\ref{eq:qeq3}) as a 5\textsuperscript{th} equation:
		\begin{equation}
			k_x (\delta Q_x^+ - \delta Q_x^-) + k_z (\delta Q_z^+ - \delta Q_z^-) = 0.
		\end{equation}
		Next we use the standard dispersion relation of magnetosonic waves. Assuming that $|k| = 1$, so that $k_z = \mathrm{cos(\theta)}$ and $k_x = 
\mathrm{sin(\theta)}$, with $\theta$ being the angle between the
background magnetic field $B_0\vec{e}_z$ and the wavevector $\vec{k}$, and that there are no background flows $V_0 = 0$, the dispersion relation is
		\begin{equation}
			\alpha B_0 \mathrm{cos(\theta)}\left(V_{A0}^2 v_{s0}^2 \mathrm{cos^2( \theta)} - \omega^2(V_{A0}^2 + v_{s0}^2) + \omega^4 \right) = 0.
			\label{eq:disprel}
		\end{equation}
		Note that we have not yet assumed any form for $\alpha$, which is not needed for isolating the magnetoacoustic solutions. If we assume a form
for $\alpha$ like in Eq.~\ref{eq:alpha}, we recover the 5\textsuperscript{th}, trivial solution of the dispersion relation, $\omega = 0$, the entropy
wave, which represents non-propagating perturbations of plasma density and temperature. The other four solutions are the up- and downward-propagating
(with respect to $\vec{e}_z$) fast and slow magnetoacoustic modes, as found also elsewhere through e.g., the velocity representation of MHD
\citep{goedbloed2004}:
		\begin{equation}
			\omega_{s,f} = \pm_\mathrm{ud} \sqrt{\frac{1}{2} \left(V_{A0}^2+ v_{s0}^2\right)} \sqrt{1\pm_\mathrm{sf} \sqrt{1-\frac{\cos ^2(\theta )
\left(4 V_{A0}^2  v_{s0}^2\right)}{\left(V_{A0}^2+ v_{s0}^2\right)^2}}}. \label{eq:slowfast}
		\end{equation}
		Here we have 4 solutions with the symbol $\pm_\mathrm{ud}$ differentiating between upward and downward-propagating waves, and the
symbol $\pm_\mathrm{sf}$ is the usual differentiation between the slow and fast magneto-acoustic waves. Recovering the magnetoacoustic solutions
demonstrates the validity of the formulation of compressible MHD equations in terms of the $Q$-variables.\par
Using the eigenvalues in terms of $\omega$ (Eq.~\ref{eq:slowfast}), the
eigenfunctions for $\vec{Q^\pm}$ can be determined for fast and slow waves from Eqs.~\ref{eq:qeq1}-\ref{eq:qeq3}. The $Q$-variables can also be 
computed directly from the velocity and
magnetic field eigenfunctions, if we assume a form for $\alpha$. Note that the definition of $\alpha$ from Eq.~\ref{eq:alpha} diverges for purely
perpendicularly-propagating ($k_z = 0$) fast waves, thus this definition is not suitable for fast waves. This uncovers a curious property of
Eqs.~\ref{eq:Lr}-\ref{eq:Lz}, in that advection (in the form of the co-moving advective derivative) is only explicitly present along the magnetic
field, leaving the definition of the phase speed in $\alpha$ only in terms of $k_z$. A straightforward remedy is then to use the full magnitude of the
wavevector instead of only the $k_z$ component in the definition of $\alpha = \omega k^{-1} B_0^{-1}$.\par

\begin{figure}[h]
			\centering
			\begin{tabular}{lr}
				\includegraphics[height=0.25\textwidth]{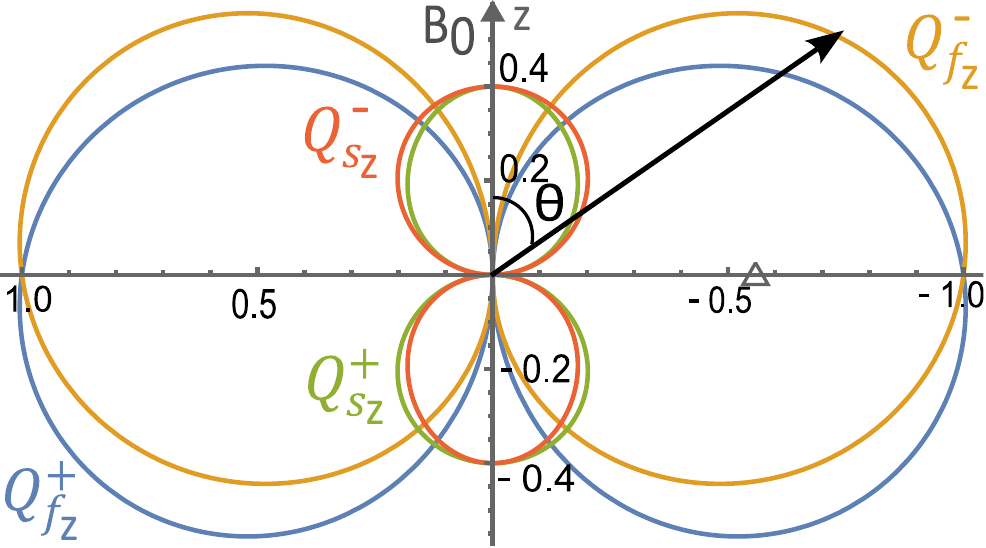}
				\hspace{0.2\textwidth}
				\includegraphics[height=0.25\textwidth]{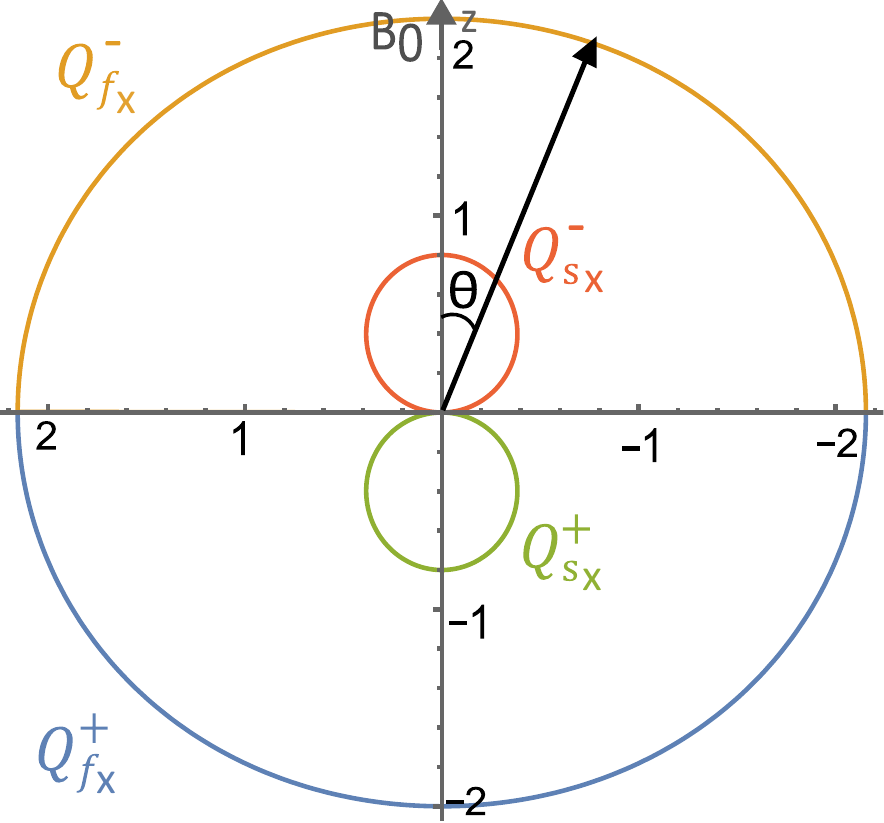}
			\end{tabular}
			\caption{Polar plots representing the $\theta$-dependence of the magnitude of parallel (left) and perpendicular (right) $Q^\pm$-variables,
for both fast and slow waves, indicated with the subscripts $f$ and $s$, respectively. The magnitudes are normalized by multiplying with the phase
speed $\omega_{s,f}/|k|$ where $|k| = 1.$ Here plasma-$\beta$ is set to 0.2. }
			\label{Qpolar}
		\end{figure}

In Figure~\ref{Qpolar} we represent the
parallel and perpendicular eigenfunctions of the $Q$-variables for fast and slow waves.
		From this figure, it is clear that only the perpendicular components are separated as a function of propagation direction with respect to the
background magnetic field. In other words, $Q^+_{s,f \perp}$ is nonzero only when $\vec{k}\cdot\vec{B_0} > 0$, and $Q^-_{s,f \perp}$ is nonzero for
$\vec{k}\cdot\vec{B_0} < 0$. The parallel components  $Q^\pm_{s,f \parallel}$, are generally both perturbed, thus based on the present form of the
$Q$-variables parallel perturbations cannot be separated into parallel and anti-parallel propagating components. The parallel components $Q^\pm_{s,f
\parallel}$ vanish only for purely parallel-propagating fast waves, which are just \alfven{} waves polarized in-plane. In the next section
(Sec.~\ref{sec:kink}), we show that this is because of the connection of $Q_\parallel$ to the magnetic pressure. \\ We conjecture that
the full separation of waves, including the component parallel to the background field, is possible by constructing a waveframe variable which
includes the total pressure or density perturbation as well in its formulation, but this will solely work in a homogeneous plasma where such a neat
separation is possible.

		\subsubsection{Kink waves}\label{sec:kink}

		In order to model kink waves, we start from Eqs.~\ref{eq:qeq1}-\ref{eq:qeq2}, written out in components. Once again, we use the same
frame of reference: the magnetic field $\vec{B}_0$ is pointing in the $z$-direction, and we also take the flow in the $z$-direction
$\vec{V}_0=V_0\vec{e_z}$. Additionally, we take the assumption of a pressureless plasma $v_\mathrm{s}=0$ and we take a density step function at $x=0$,
with a constant density $\rho_\mathrm{L}$ ($\rho_\mathrm{R}$) on the left (right) side of the interface. In each half space, the waves may be Fourier
analysed in $y$, $z$ and $t$, putting every quantity proportional to $\exp{(i k_z z-i\omega t)}$, where we have once again considered $k_y\equiv 0$ as
in Subsect.~\ref{sec:linear}. The resulting equations will be just like equations \ref{eq:Lr}-\ref{eq:Lz}, except that the terms with $k_x$ will be
replaced by a derivative $d/dx$. In what follows, we ignore Eq.~\ref{eq:Ly}, because we will not concentrate on the \alfven{} waves, but rather on the
kink waves, which are solely polarised in the $x,z$-directions for $k_y=0$.\par

		Following the earlier strategy, we take (e.g.) $\omega^+_\mathrm{L,R}=0$ to find the upward propagating kink waves. This immediately implies a
connection
		\begin{equation}
			V_\mathrm{L}+\alpha_\mathrm{L}B_\mathrm{L}=V_\mathrm{R}+\alpha_\mathrm{R}B_\mathrm{R} \label{eq:matchom}
		\end{equation}
		between $\alpha_\mathrm{R,L}$. Each quantity in this equation is the corresponding background quantity in the left half space or right half
space respectively for subscripts L and R. With this assumption, we then have the following set of equations:
		\begin{align}
			0 &= -\frac{1}{2} \frac{d}{dx} (3 \delta Q^+_x-\delta Q^-_x)-\frac{1}{2} k_z (3\delta Q^+_z-\delta Q^-_z) + \alpha B_0k_z \delta R,\\*
			0 &=  - \frac{1}{4}\left(1-\frac{\Delta \alpha^2}{\alpha^2}\right) (Q_0^+-Q_0^-) \frac{d}{dx} (\delta Q^+_z-\delta Q^-_z) \nonumber \\ & 
- \frac{1}{4}
\frac{\Delta \alpha^2}{\alpha^2} (Q_0^+-Q_0^-)ik_z (\delta Q^+_x-\delta Q^-_x),\label{eq:ompx}\\*
			- ik_z(Q_0^+-Q_0^-)\delta Q^+_x &= - \frac{1}{4}\left(1-\frac{\Delta \alpha^2}{\alpha^2}\right) (Q_0^+-Q_0^-) \frac{d}{dx} (\delta 
Q^+_z-\delta
Q^-_z) \nonumber \\ &- \frac{1}{4} \frac{\Delta \alpha^2}{\alpha^2} (Q_0^+-Q_0^-)ik_z (\delta Q^+_x-\delta Q^-_x), \label{eq:ommx}\\*
			-\frac{Q_0^+-Q_0^-}{4}\left[\frac{d}{dx}\right. & \left.(\delta Q_x^++\delta Q_x^-)  +ik_z(\delta Q_z^++\delta Q_z^-)\right] = \nonumber 
\\ & -
\frac{1}{4}\left(1-\frac{\Delta \alpha^2}{\alpha^2}\right) (Q_0^+-Q_0^-) ik_z (\delta Q^+_z-\delta Q^-_z) \nonumber \\* & + \frac{1}{4}
\frac{\Delta \alpha^2}{\alpha^2} (Q_0^+-Q_0^-) \frac{d}{dx} (\delta Q^+_x-\delta Q^-_x), \label{eq:ompz}\\*
			- ik_z(Q_0^+-Q_0^-)\delta Q^+_z & +\frac{Q_0^+-Q_0^-}{4}\left[\frac{d}{dx}(\delta Q_x^++\delta Q_x^-) +ik_z(\delta Q_z^++\delta
Q_z^-)\right] \nonumber \\* & = - \frac{1}{4}\left(1-\frac{\Delta \alpha^2}{\alpha^2}\right) (Q_0^+-Q_0^-) ik_z (\delta Q^+_z-\delta Q^-_z)  \nonumber 
\\ &+ 
\frac{1}{4}
\frac{\Delta \alpha^2}{\alpha^2} (Q_0^+-Q_0^-) \frac{d}{dx} (\delta Q^+_x-\delta Q^-_x), \label{eq:ommz}
		\end{align}
		in which all quantities are subscripted with R and L respectively for each half space.
		Combining Eq.~\ref{eq:ommx} and Eq.~\ref{eq:ompx} isolates $\delta Q_x^+$ as
		\begin{equation}
			ik_z(Q_0^+-Q_0^-)\delta Q^+_x=0,
		\end{equation}
		showing that the kink wave is uniquely described by $\delta Q_x^-$ only, because $\delta Q_x^+$ is 0 if $k_z\neq 0$ and $B_0\neq 0$. If we 
find a value for
$\alpha_\mathrm{R,L}$ and $\omega$, then the kink wave is written with only one of $\delta Q_x^\pm$, as was the intention of the $Q$-variables for 
separating
upward and downward propagating waves. Similarly, from the combination of Eq.~\ref{eq:ompz} and Eq.~\ref{eq:ommz} (and using $\delta Q_x^+=0$), we 
obtain
		\begin{equation}
			\frac{d}{dx} \delta Q_x^- - ik_z (\delta Q^+_z-\delta Q^-_z)=0. \label{eq:ompdiv}
		\end{equation}
		Thus, we obtain a set of equations describing the kink waves (or any other wave under these assumptions) from Eq.~\ref{eq:ompx} and
Eq.~\ref{eq:ompdiv}
		\begin{align}
			0&=\frac{d}{dx} \delta Q_x^- - ik_z (\delta Q^+_z-\delta Q^-_z),\\*
			0 &=  \left(1-\frac{\Delta \alpha^2}{\alpha^2}\right)  \frac{d}{dx} (\delta Q^+_z-\delta Q^-_z) - \frac{\Delta \alpha^2}{\alpha^2} ik_z 
\delta Q^-_x.
		\end{align}
		Introducing a new variable $\Pi=\delta Q^+_z-\delta Q^-_z$, we obtain the set
		\begin{align}
			0&=\frac{d}{dx} \delta Q_x^- - ik_z \Pi,\label{eq:qxpi}\\*
			0 &=  \left(1-\frac{\Delta \alpha^2}{\alpha^2}\right)  \frac{d\Pi}{dx}  - \frac{\Delta \alpha^2}{\alpha^2} ik_z \delta Q^-_x.
		\end{align}
		This set is reminiscent of the coupled differential equations between perturbed total pressure and displacement that other works have found
for the description of kink waves \citep{appert1974,goossens1992,ismayilli2022}, which have a strong correspondence to the currently modelled 
surface \alfven{} waves \citep{goossens2012}.\par

		Since each quantity is constant in each half space, we can substitute one of the equations in the other. Then, we obtain a single 2nd order
differential equation:
		\begin{equation}
			\frac{d^2}{dx^2}\delta Q^-_x +k_z^2 \left(\frac{\Delta \alpha^2}{\alpha^2-\Delta \alpha^2}\right) \delta Q^-_x=0.
		\end{equation}
		In the left and right half space, we consider respectively the solution
		\begin{equation}
			\delta Q^-_{x,\mathrm{L}}=A_\mathrm{L} \exp{(\kappa_\mathrm{L} x)},\quad \delta Q^-_{x,\mathrm{R}}=A_\mathrm{R} \exp{(-\kappa_\mathrm{R}
x)},
		\end{equation}
		where
		\begin{equation}
			\kappa^2=k_z^2\left\vert \frac{\Delta \alpha^2}{\Delta \alpha^2-\alpha^2}\right\vert.
		\end{equation}
		The solution for $\Pi$ can be calculated from Eq.~\ref{eq:qxpi}.\par

		Next, we need to apply boundary conditions at $x=0$. Namely, we take (as usual)
		\begin{align}
			0&=[v_x],\\*
			0&=[P'],
		\end{align}
		where $P'=B_0b_z/\mu$ is the perturbed total pressure and the square brackets are differences between the left and right of the interface.
Note that the extra term in the Lagrangian pressure perturbation is 0 in the linear regime, since the magnetic pressure is uniform in each
half space. Translated to our variables, these boundary conditions are
		\begin{align}
			0&=[Q_x^-],\\*
			0&=\left[\frac{B_0\Pi}{\alpha}\right].
		\end{align}
		The first condition states that $A_\mathrm{L}=A_\mathrm{R}$, while the second condition results in the dispersion relation:
		\begin{equation}
			-\frac{\kappa_\mathrm{L}B_\mathrm{L}}{\alpha_\mathrm{L}}=\frac{\kappa_\mathrm{R}B_\mathrm{R}}{\alpha_\mathrm{R}}.
		\end{equation}
		Squaring this relation, and inserting the expression for $\kappa^2=k_z^2|\mu\rho\alpha^2-1|$, we obtain
		\begin{equation}
			\left(\mu\rho_\mathrm{L}-\frac{1}{\alpha_L^2}\right)B_\mathrm{L}^2=-\left(\mu\rho_\mathrm{R}-\frac{1}{\alpha_R^2}\right)B_\mathrm{R}^2,
		\end{equation}
		where we have used the fact that the absolute values in $\kappa^2$ take a different sign on either side of the interface. Solving this
equation in conjunction with Eq.~\ref{eq:matchom} (and considering $V_0=0$ for simplicity), we obtain finally the allowed values for
$\alpha$:
		\begin{equation}
			\alpha
B_0=\alpha_\mathrm{L}B_\mathrm{L}=\alpha_\mathrm{R}B_\mathrm{R}=\sqrt{\frac{B_\mathrm{R}^4+B_\mathrm{L}^4}{\mu\left(\rho_\mathrm{R}B_\mathrm{R}^2+ 
\rho_\mathrm{L}B_\mathrm{L}^2\right)}}, \qquad
\omega=k_z\sqrt{\frac{B_\mathrm{R}^4+B_\mathrm{L}^4}{\mu\left(\rho_\mathrm{R}B_\mathrm{R}^2+\rho_\mathrm{L}B_\mathrm{L}^2\right)}}.
		\end{equation}
		Given that $B_\mathrm{L}=B_\mathrm{R}=B_0$ for a pressureless plasma, these equations reduce to
		\begin{equation}
			\alpha =\alpha_\mathrm{L}=\alpha_\mathrm{R}=\sqrt{\frac{2}{\mu\left(\rho_\mathrm{R}+\rho_\mathrm{L}\right)}},\qquad
\omega=k_z\sqrt{\frac{2B_0^2}{\mu\left(\rho_\mathrm{R}+\rho_\mathrm{L}\right)}}, \label{eq:alphakink}
		\end{equation}
		as is well known from other works.

		\subsubsection{General waves in field-aligned flows}
		Now we will prove explicitly that the proper choice of $\alpha$ splits the $Q$-variable between wave modes of propagation directions. We 
follow the derivation of \citet{magyar2019} and their Eq.~19. In this subsection, we consider the general configuration with a magnetic 
field pointing in the $z$-direction, but still dependent on $x$ and $y$. Moreover, we also take the background flow along the magnetic field.
		\begin{equation}
		    \vec{B}_0=B_0(x,y)\vec{e}_z, \vec{V}_0=V_0(x,y)\vec{e}_z.
		\end{equation}
		Let us now consider the linearised induction equation
		\begin{equation}
		    \frac{\partial \vec{b}}{\partial t}=\nabla\times\left((\vec{V_0}+\vec{v})\times \vec{B}_0\right),
		\end{equation}
		of which we will only consider the perpendicular component. We can reduce this induction equation with vector identities to 
		\begin{equation}
		    \frac{\partial \vec{b}_\perp}{\partial t}=\vec{B}_0\cdot\nabla\vec{v}_\perp - \vec{V}_0\cdot\nabla 
\vec{b}_\perp.
		\end{equation}
		Here we have naturally used that $\nabla\cdot \vec{B}_0=\nabla \cdot \vec{b}=0$, but we have also used $\nabla \cdot \vec{V}_0=0$ because 
$\vec{V}_0$ only has a $z$-component that does not depend on $z$. Using Fourier analysis for the ignorable coordinates $z$ and $t$, we then have 
		\begin{equation} 
		    -i\omega \vec{b}_\perp = ik_z(B_0 \vec{v}_\perp - V_0 \vec{b}_\perp)
		\end{equation}
		Using Eq.~\ref{eq:B} and \ref{eq:V}, we then have 
		\begin{equation}
		    (\omega-k_zV_0+k_z\alpha B_0) \vec{\delta Q}^+_\perp = (\omega - k_z V_0-k_z\alpha B_0) \vec{\delta Q}^-_\perp.
		\end{equation}
		This equation shows that the correct choice of $\alpha$ indeed splits a wave mode with a specific $\omega$ and $k_z$ between different 
$\vec{\delta Q}_\perp$ components. Using the $Q$-variable terminology, the equation is more elegantly written as 
		\begin{equation}
		    (\omega - \vec{k}\cdot \vec{Q}_0^-)\vec{\delta Q}^+_\perp = (\omega - \vec{k}\cdot \vec{Q}_0^+)\vec{\delta Q}^-_\perp.
		\end{equation}
		This equation states that a wave with phase speed $\vec{Q}_0^\pm$ has the associated wave only present in $\vec{\delta Q}_\perp^\mp$, with 
the other $Q$-variable $\vec{\delta Q}_\perp^\pm =0$.

		\subsection{Splitting the equations for different wave modes}
		The linearised $Q$-equations (Eqs.~\ref{eq:qeq1}-\ref{eq:qeq2}) and their component versions (Eqs.~\ref{eq:Lr}-\ref{eq:Lz}) show that the
operator in the RHS of these equations is a linear operator, and yields a vector proportional to its input plane wave solution with dependence
$\exp{(i\vec{k}\cdot\vec{x}-i\omega t)}$. Moreover, we understand that the wave vector $\vec{k}$ and $\omega$ must satisfy the dispersion relation.

		In a future work, we want to construct models for the solar atmosphere, which are driven by different wave modes. In our upcoming models, we 
want to take a step back from the linear approach,
and once again use the full operator. The plan is to use a WKB approach as detailed in \citet{marsch1989, tu1993, vanderholst2014}. Such a model with 
only \alfven{} wave drivers is called AWSOM (\alfven{} Wave driven Solar Model) models \citep{vanderholst2014}. We want to extend this model by also 
including the kink waves, their self-interaction and damping, leading to a model named UAWSOM (Uniturbulence and \alfven{} Wave 
driven Solar Model). Thus we are looking forward to taking
		\begin{equation}
			\vec{Q}^\pm=\vec{Q}_0^\pm+\vec{\delta Q}^\pm_\mathrm{k}+\vec{\delta Q}^\pm_\mathrm{A}, \label{eq:deltaq}
		\end{equation}
		where $\vec{Q}_0^\pm$ stands for the slowly varying background, $\vec{\delta Q}^\pm_\mathrm{k}$ is the contribution of the (respectively up-
and downward propagating) kink waves, and $\vec{\delta Q}^\pm_\mathrm{A}$ the contribution from the (up- and downward) \alfven{} waves, as
classically used in AWSOM type models \citep{vanderholst2014, evans2012, reville2020}. In the employed WKB approximation, we consider a background
$\vec{Q}_0^\pm$ slowly varying along $\vec{B}_0$ and in time. We thus consider the dominant Fourier components of $\vec{Q}_0^\pm$ to be with
wavelengths (or scale heights, if you wish) much larger than wavelengths of the kink and \alfven{} waves, and periods (or time scales of variation, if
you like) much larger than the periods of the kink and \alfven{} waves.

		Let us first only consider the linearised version of Eqs.~\ref{eq:qmhd}, \ref{eq:clos2}, and call its associated operator ${\cal L}_\alpha$ 
acting on the
eigenvector to be $\vec{\cal U}$ (consisting of $\vec{Q}^\pm$ and $\rho$):
		\begin{equation}
			{\cal L}_\alpha \vec{\cal U} = 0.
		\end{equation}
		We realise that the linear operators in the LHS and RHS will just split out over the different contributions $\vec{\delta Q}^\pm_\mathrm{k}$
and $\vec{\delta Q}^\pm_\mathrm{A}$, because of the linear character of the operators. Each term for the kink wave in the equation will have a
dependence $\exp{(i k_{z,\mathrm{k}}z-i\omega_\mathrm{k} t)}$, and likewise the terms for the \alfven{} waves will have a dependence of
$\exp{(ik_{z,\mathrm{A}}z-i\omega_\mathrm{A} t)}$, where the pairs $(\omega_\mathrm{k}, k_{z,\mathrm{k}})$ and $(\omega_\mathrm{A}, k_{z,\mathrm{A}})$
satisfy their respective dispersion relation for a (different!) driving frequency $\omega_\mathrm{k}$ or $\omega_\mathrm{A}$ that finds its origin in
the photospheric convective motions or p-modes \citep{morton2019}. By using a Fourier transform of the linearised $Q$-equations, we then obtain a
separated set of equations for each contribution:
		\begin{align}
			{\cal L}_\alpha \vec{\cal U}_\mathrm{k} & = 0, \label{eq:Lkink}\\
			{\cal L}_\alpha \vec{\cal U}_\mathrm{A} & = 0, \label{eq:Lalfven}\\
			{\cal L}_\alpha \vec{\cal U}_0 & = 0.
		\end{align}
		Here the last equation for the equilibrium is in the WKB approximation an integration of the Fourier components $\omega$ smaller than the
smallest wave frequency:
		\begin{equation}
			\omega < \min{\left\lbrace \omega_\mathrm{A}, \omega_\mathrm{k}\right\rbrace},
		\end{equation}
		which thus represents the slow evolution of the background. It is irrelevant for this last equation for the equilibrium which $\alpha$ value
is chosen or used, because the equations are more conveniently written in terms of the classical MHD variables.

		The key point to realise in Eqs.~\ref{eq:Lkink}-\ref{eq:Lalfven} is that they are still valid for any possible $\alpha$ that you prefer.
Moreover, they are clearly independent, and thus $\alpha$ may be chosen freely for both separately! Thus, for Eq.~\ref{eq:Lalfven}, we use the choice
of $\alpha=1/\sqrt{\mu\rho}$ reverting to the classical equation of \citet{vanderholst2014}. However, for the kink waves (Eq.~\ref{eq:Lkink}), we
make the choice of the appropriate $\alpha$, as found in Eq.~\ref{eq:alphakink}. That then allows to formulate the appropriate equations for upward
and downward propagating kink waves, separating out their contributions.

		If we assume that the non-linearity and field-aligned inhomogeneity is sufficiently weak, we can consider the re-inclusion of the non-linear 
terms in Eq.~\ref{eq:qmhd} and \ref{eq:clos2}. They will be of the form $\vec{\delta Q}^\pm_\mathrm{k}
\cdot \nabla \vec{\delta Q}^\pm_\mathrm{k}$ and $\vec{\delta Q}^\pm_\mathrm{A} \cdot \nabla \vec{\delta Q}^\mp_\mathrm{A}$, and also include
cross-terms between $\vec{\delta Q}_\mathrm{A}$ and $\vec{\delta Q}_\mathrm{k}$. Using the same Fourier argument as before, we should realise that the
cross-terms will have no net contribution to the equations Eqs.~\ref{eq:Lkink}-\ref{eq:Lalfven} when integrated over a longer time \citep[this seems,
however, in contradiction with the numerical experiments of][]{guo2019b}. The other terms will contain the classical interaction of counterpropagating
waves in \alfven{} wave turbulence \citep{iroshnikov1964, kraichnan1967}, acting as a net sink in the equations Eqs.~\ref{eq:Lkink}-\ref{eq:Lalfven},
but added as a source term in the equilibrium equations as in \citet{marsch1989, tu1993, vanderholst2014, evans2012, reville2020}. The terms in
$\vec{\delta Q}^\pm_\mathrm{k} \cdot \nabla \vec{\delta Q}^\pm_\mathrm{k}$ model the damping of the kink wave due to uniturbulence
\citep{magyar2017,magyar2019} due to its self-deformation. In \citet{vd2020} it was found that this term also leads to a net contribution when
averaged over longer times, similar to the \alfven{} wave cascade. This extra contribution also acts as a sink in the kink wave evolution equation
(Eq.~\ref{eq:Lkink}), and is added as an extra heating and pressure term in background MHD equations, just like the \alfven{} wave cascade in the
AWSOM model.

	\section{Conclusions}
	In this paper, we have started from the success of the \elsasser{} variables in describing and separating upward and downward propagating
\alfven{} waves. With the earlier realisation that any other wave than an \alfven{} wave necessarily has both \elsasser{} components
\citep{magyar2019}, we have realised that the \elsasser{} variables need generalisation to other waves as well.

To fill this need, we have proposed the $Q$-variables given by
\begin{equation*}
    \vec{Q}^\pm=\vec{V}\pm \alpha\vec{B},
\end{equation*}
with a parameter $\alpha$ that we have proven to be proportional to the phase speed of the wave. The value of $\alpha$ is dependent on the type of
wave and equilibrium
parameters through the dispersion relation. We have rewritten the MHD equations in these $Q$-variables, following the lead of \citet{marsch1987}.

In the next section of the paper, we have shown that (1) the modelling of \alfven{} waves reverts back to the classical \elsasser{} variables, (2)
that slow and fast waves have also the perpendicular component of $\vec{Q}^\pm$ split between upward and downward propagating waves, and (3) that
surface
\alfven{} waves in a non-uniform plasma can also be described by the $Q$-variables, separating out upward and downward propagating waves. This shows
that the generalisation of the \elsasser{} variables as we set out to do, has been successful. Indeed, going beyond the \elsasser{} description, with
the current $Q$-variables, we can separate upward and downward propagating waves of many different types, including waves in inhomogeneous plasmas.

The significance of these $Q$-variables is in enabling a more general approach to the \alfven{} wave driven solar wind models
\citep[e.g.][]{vanderholst2014}. These models encapsulate in a 1D way the additional heating by \alfven{} waves \citep[see][for a
review]{cranmer2015}. Thanks to this new development of the $Q$-variables, it will be possible to construct new solar wind models that
also include wave driving by other wave modes. In particular, we have laid the mathematical groundwork for the creation of the UAWSOM
model, which also incorporates the propagation of kink waves on inhomogeneous structures, such as plumes. Kink waves have been ubiquitously observed
in the solar corona \citep{tomczyk2007,nechaeva2019} and possibly deliver significant energy input in coronal loops \citep{lim2023} and plumes
\citep{thurgood2014}. These kink waves self-interact non-linearly
and show uniturbulence \citep{magyar2019}. This potentially leads to extra heating in the solar wind model, possibly resolving current shortcomings
of the AWSOM model which
underperforms in open field regions \citep{verdini2010,vanderholst2014,verdini2019,vanballegooijen2016,vanballegooijen2017}. This potential extra
heating by kink
waves will be the subject of a future publication, in which we will derive the governing equations for the UAWSOM model, based on the current
$Q$-variables. These will incorporate the evolution equations of the wave energy density. Moreover, but more speculatively, this formalism could be 
useful in deriving the effect of the parametric instability on the solar wind driving with \alfven{}
waves \citep{shoda2019}.

Furthermore, the adoption of these new $Q$-variables allows the exploration Solar Orbiter or Parker Solar Probe data, in regimes which are not highly 
Alfv\'enic. In particular, data series of low Alfv\'enicity could be re-analysed with the $Q$-variables to expose other wave modes in these regimes.

\section*{Acknowledgements}
This research was supported by the International Space Science Institute (ISSI) in Bern, through ISSI International Team project \#560: ``Turbulence
at the Edge of the Solar Corona: Constraining Available Theories Using the Latest Parker Solar Probe Measurements''.

\section*{Funding}
TVD was supported by the European Research Council (ERC) under the European Union's Horizon 2020 research and innovation programme (grant
agreement No 724326), the C1 grant TRACEspace of Internal Funds KU Leuven, and a Senior Research Project (G088021N) of the FWO Vlaanderen. 
Furthermore, TVD received financial support from the Flemish Government under the long-term structural Methusalem funding program, project SOUL: 
Stellar evolution in full glory, grant METH/24/012 at KU Leuven. NM
acknowledges Research Foundation – Flanders (FWO Vlaanderen) for their support through a Postdoctoral Fellowship.
MVS acknowledge support from the French Research Agency grant ANR STORMGENESIS \#ANR-22-CE31-0013-01.

\section*{Declaration of Interests} 
The authors report no conflict of interest.

\section*{Author ORCID}
\noindent T. Van Doorsselaere, https://orcid.org/0000-0001-9628-4113; \\ M. V. Sieyra, https://orcid.org/0000-0002-1536-8508

\section*{Author contributions}
TVD derived the theory, NM made the numerical solutions, all contributed in discussions during the research, all contributed in writing and editing 
the manuscript. 

\def\aj{AJ}%
\def\araa{ARA\&A}%
\def\apj{ApJ}%
\def\apjl{ApJ}%
\def\apjs{ApJS}%
\def\ao{Appl.~Opt.}%
\def\apss{Ap\&SS}%
\def\aap{A\&A}%
\def\aapr{A\&A~Rev.}%
\def\aaps{A\&AS}%
\def\azh{AZh}%
\def\baas{BAAS}%
\def\jrasc{JRASC}%
\def\memras{MmRAS}%
\def\mnras{MNRAS}%
\def\pra{Phys.~Rev.~A}%
\def\prb{Phys.~Rev.~B}%
\def\prc{Phys.~Rev.~C}%
\def\prd{Phys.~Rev.~D}%
\def\pre{Phys.~Rev.~E}%
\def\prl{Phys.~Rev.~Lett.}%
\def\pasp{PASP}%
\def\pasj{PASJ}%
\def\qjras{QJRAS}%
\def\skytel{S\&T}%
\def\solphys{Sol.~Phys.}%
\def\sovast{Soviet~Ast.}%
\def\ssr{Space~Sci.~Rev.}%
\def\zap{ZAp}%
\def\nat{Nature}%
\def\iaucirc{IAU~Circ.}%
\def\aplett{Astrophys.~Lett.}%
\def\apspr{Astrophys.~Space~Phys.~Res.}%
\def\bain{Bull.~Astron.~Inst.~Netherlands}%
\def\fcp{Fund.~Cosmic~Phys.}%
\def\gca{Geochim.~Cosmochim.~Acta}%
\def\grl{Geophys.~Res.~Lett.}%
\def\jcp{J.~Chem.~Phys.}%
\def\jgr{J.~Geophys.~Res.}%
\def\jqsrt{J.~Quant.~Spec.~Radiat.~Transf.}%
\def\memsai{Mem.~Soc.~Astron.~Italiana}%
\def\nphysa{Nucl.~Phys.~A}%
\def\physrep{Phys.~Rep.}%
\def\physscr{Phys.~Scr}%
\def\planss{Planet.~Space~Sci.}%
\def\procspie{Proc.~SPIE}%

\bibliography{refs}
\bibliographystyle{jpp}

\end{document}